\documentclass[aps,prl,showpacs,twocolumn,superscriptaddress]{revtex4}
\usepackage{graphicx}
\usepackage{amsmath}
\usepackage{amssymb}
\bibstyle{apsrev.bib}

\begin{document}

\title{When can Fokker-Planck Equation
describe anomalous or chaotic transport ?}

\author{D.F. Escande}

\address{UMR 6633 CNRS-Universit\'{e} de Provence, Marseille, France}

\address{Consorzio RFX, Associazione EURATOM-ENEA sulla fusione,
Padova, Italy}

\author{F. Sattin}

\address{Consorzio RFX, Associazione EURATOM-ENEA sulla fusione,
Padova, Italy}

\date{\today}

\begin{abstract}
The Fokker-Planck Equation, applied to transport processes in fusion plasmas,
 can model several anomalous features, including uphill transport,
scaling of confinement time with system size, and convective
propagation of externally induced perturbations. It can be justified
for generic particle transport provided that there is enough randomness in the
Hamiltonian describing the dynamics. Then, except for 1
degree-of-freedom, the two transport coefficients are largely
independent. Depending on the statistics of interest, the same dynamical system
may be found
diffusive or dominated by its L\'{e}vy flights.

\end{abstract}


\maketitle

The Fokker-Planck Equation (FPE) is a basic model for the description
of transport processes in several scientific fields. In one dimension
it reads
\begin{equation}
\partial_t n(x,t) =  - \partial_x \left( {V(x)n} \right) +
\partial_x^2 \left( {D(x)n} \right)
\label{eq1}
\end{equation}
where $n$ is the density for a generic scalar quantity, $V$ the dynamic friction,
and $D$
the
diffusion coefficient. In particular, FPE backs up the diffusion-convection picture
of
anomalous transport in magnetized
thermonuclear fusion plasmas, where $V$ is often referred to as a pinch velocity.
This transport stems from turbulence triggered by
density and temperature gradients through convective instabilities. Though very
popular, the
drift-diffusive
picture underlying FPE breaks down in some cases. This was
proved, e.g., for the transport  of tracer
particles suddenly released in pressure-gradient-driven
turbulence, which exhibits strongly non gaussian features \cite{ref1}.
This fact triggered a series of
studies where transport was described in terms of continuous random
walks with L\'{e}vy jumps, and of fractional
diffusion models (see \cite{ref2} and references therein). This sets
the issue: when is FPE relevant for anomalous or chaotic transport, when is it
not?

Our discussion has many facets, but goes through the following main steps
: (\textit{i}) FPE with a source term
can model phenomena commonly labeled as
\textquotedblleft{}anomalous\textquotedblright{} in fusion plasmas: uphill
transport
\cite{ref3}, anomalous scaling of confinement time with system size \cite{ref4},
and non diffusive propagation of externally induced perturbations
\cite{ref5}. (\textit{ii}) If the source is narrower than the mean random step of
the true
dynamics, FPE fails, while the correct Chapman-Kolmogorov equation reveals
the existence
of spatial features of the spreading quantity that are not related to the transport
coefficients.
(\textit{iii}) For general 1 degree-of-freedom (dof)
  Hamiltonian
dynamics, the constraint \textit{V = dD/dx}  holds, as originally
derived by Landau. This
constraint vanishes for higher dimensional dynamics, incorporating for
instance particle-turbulence self-consistency, or the full 3-dimensional motion of
particles. (\textit{iv}) FPE can be justified for particle transport provided
that there is enough randomness in the Hamiltonian describing the
dynamics: e.g., when it includes many waves with random phases.
This may work even whenever the dynamics of individual particles
exhibit strong trapping motion. However chaos is not enough to justify FPE.
(\textit{v}) The diffusion may be quasilinear (QL)
or not, depending on the so-called Kubo number  $K $, which scales like the
correlation time
\cite{ref6,ref7}. For  $K $ small, QL diffusion is due
to the locality in velocity of wave-particle resonance. Large
 $K $ diffusion is due to the locality of trapping in
phase-space.

Our discussion aims at providing insight in transport mechanisms, rationales for
confinement scaling laws, and tools for experimental data analysis. It involves
several building blocks: some are provided by the available literature, while
a few are elaborated here. Though written for the fusion community, a large part
of our discussion is of direct relevance to chaotic transport, and in particular to
Lagrangian dynamics in incompressible turbulence.

\textit{What FPE can do.}
 The case where $ V $ and
 $D$ are constant in Eq. (\ref{eq1}) is often considered in fusion data
analysis. An initially Dirac-like perturbation travels with velocity
\textit{V}, while diffusing with coefficient $D$. For large enough systems, $L
\gg D/V$, the perturbation actually travels ballistically.
Thus FPE can model the non diffusive propagation of  induced
perturbations in a fusion machine \cite{ref5}. The scaling of confinement time
$\tau_\text{conf}$
with system size $L$, goes from diffusive
($\tau_\text{conf} \propto L^2$, $L \ll D/V$) to ballistic
($\tau_\text{conf} \propto L^1$, $L \gg D/V$). Thus FPE may account for
anomalous scaling of confinement time with system size.
It is even more so when $V$ and $D$ depend on $x$ \cite{ref4}, or when their 
possible
dependence on spatial gradients driving turbulence is accounted for. Whatever
$V$ and $D$ be,
FPE may be written as $\partial_t n = - \partial_x \Gamma $, with the flux
$\Gamma = V n - \partial_x (D n) $ . For $D$ constant, if there are no
sources in the central part of the plasma, for symmetry reasons
$\Gamma =0$, which yields $\partial_x n/n = V/D $: if $sign(x)V < 0$, where
$x=0$
corresponds to the plasma center,
the probability piles up toward small $|x|$'s, leading to an equilibrium
distribution of
\textquotedblleft{}uphill transport\textquotedblright{} type \cite{ref3}. However,
the piling up may
result as well from $V=0$, and $D(x)$ growing with $|x|$ since
$ n(x)\propto 1/D(x)$. This is a caveat for data analysis: a broad family of
$(V,D)$ profiles
can model the same experimental data.  Let us notice that the ability of FPE to
describe
anomalous transport is shared with models giving non gaussian features (see
\cite{ref2} and
references therein).

\textit{ Transport in presence of a source}. The classical derivation of FPE from
the
Chapman-Kolmogorov equation
\begin{equation}
\partial_t n(x,t) = \int P(x,x') {n(x',t) \over \tau(x') }dx' - {n(x,t) \over \tau(x)} +
S(x),
\label{eq3}
\end{equation}
where $S(x)$ is a localized source term and $\tau(x)$ is a waiting time, assumed
to be 1 for
the moment, assumes the typical width $\sigma_P$ of $P$ in $x-x'$   to be the
smallest spatial
scale of interest. Therefore the width $\sigma_S$ of $S$ must be much larger
than
$\sigma_P$ for FPE to describe correctly $n(x,t)$ within the source domain;
otherwise the source
pumps strong gradients in $n(x,t)$, and the Taylor expansion leading to FPE
breaks down
close to it. Assume $P(x,x')$ to be a function of $x-x'$ only. Then Eq. (\ref{eq3})
shows that for scales smaller than $\sigma_P$, the Fourier transform of a
stationary $n(x)$ is
almost equal to that of $S(x)$. These are the scales relevant to describe the
profile close to the
source. Therefore, $n(x)$ displays a bump similar to that of $S(x)$ that is not due
to specific
properties of the FPE transport coefficients: a further caveat for data analysis.

\textit{Paradigm model}.
Consider tracer transport in 2-dimensional (2D) incompressible
turbulence, or transport in magnetized plasmas induced by
electrostatic turbulence in the guiding
center approximation.
A paradigm for such a transport is the equation of motion
\begin{equation}
\frac{{d{\bf{X}}}}{{dt}} =  - \nabla \Phi({\bf{X}},t) \times {\bf{\hat
x}}_3  \equiv {\bf{v}}
\label{eq2}
\end{equation}
with  ${\bf{X}} = (x_1 ,x_2)$ the particle position (perpendicular to the magnetic
field, for the plasma case)
 and $\Phi$ the flow function or the appropriately normalized
electrostatic potential. Model (\ref{eq2}) applies to transport due to magnetic
chaos as
well. Indeed, for an almost straight magnetic field,
$\Phi$ may be replaced by $\Phi' = \Phi - v_{||} A_{||}$, where both the
electrostatic potential  $\Phi{} $ and the
parallel vector potential $A_{||}$
are computed at the guiding center position, and $v_{||}$ is the parallel velocity
\cite{ref6}.
Therefore Eq. (\ref{eq2}) describes the canonical equations for the
guiding center dynamics in a mixture of electrostatic and magnetostatic fields,
 ruled by Hamiltonian
 $\Phi(\bf{X},\rm{t})$, with conjugate variables $x_1$ and $x_2$.

\textit{Constraint on FPE due to the dimensionality}. Consider model (\ref{eq2})
where
$\Phi$ has a bounded support, or is periodic in the $x_i$'s with an elementary
periodicity cell $R$: motion is, or can be considered as, located within a bounded
domain  $R $ of phase
space.  Because of conservation of the area of a phase
space element during motion, an initially uniform particle density $n$ in  $R$
must remain uniform for later times. Now, assume FPE  describes
the evolution of $n(x)$  ($x$ one
of the two conjugate variables). Since $n$ is stationary, the
flux  $\Gamma$ must be a constant in $R$, and must vanish, since it does on the
boundary
of $R$. Since $\Gamma = V n - \partial_x (D n) $, this requires $V = dD/dx$.
 This is the fundamental reason for this
constraint, first derived by Landau \cite{ref8} for a stochastic but non chaotic 
Hamiltonian
dynamics,  since his derivation uses Taylor
expansions of the orbit in time. The Landau constraint (LC) was recovered
more recently for the chaotic motion of particles in a prescribed set
of Langmuir waves \cite{ref10}. If $V = dD/dx$, then $ \Gamma = - D dn/dx$. In
absence of
sources
in the plasma core, $\Gamma{} = 0$ implies
$dn/dx = 0$ in this domain, which rules out uphill transport. \\
Adding more dimensions to the previous Hamiltonian dynamics (for
instance the parallel motion in addition to the
\textbf{E}$\times$\textbf{B} drift) leads in
general to a breakdown of LC. Indeed a Kolmogorov-Arnold-Moser torus is no 
longer able to separate the chaotic domain into disconnected sets (Arnold 
diffusion), and therefore there are no longer boundaries $\Gamma$  must vanish 
onto.
 For instance, LC no longer
holds for the self-consistent motion of particles in a set of Langmuir
waves (here  $x$ is the particle velocity): $V$ is equal to
$dD/dx$ plus a
drag force due to the
spontaneous emission of waves by particles \cite{ref10}. The analogy of
Langmuir
wave-electron interaction with the toroidal Alfven eigenmode-fast ion one, where 
$x$
is the
radial position
\cite{ref11}, shows that a similar effect may hold for a fusion machine. This is a
particular instance revealing that the true Hamiltonian dynamics of
particles in fusion machines has more than 1 dof. In general LC cannot hold then:
uphill
transport and a central finite
density gradient are possible.

\begin{figure}

\includegraphics[width=60mm, height=55mm]{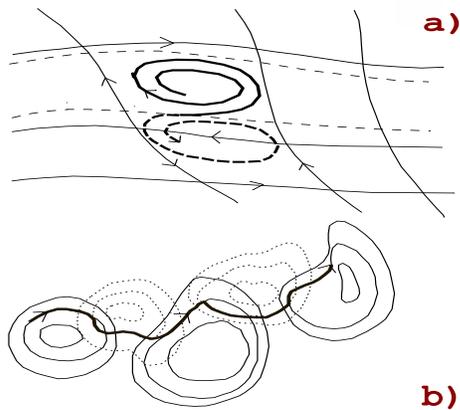}
\caption{(a) $ K \gg 1$: as the net of intersecting separatrices evolves in time 
(from solid to dashed curves), the lower domain enlarges at the expense of the 
upper one. Hence a trajectory may go from the upper to the lower one (thick to 
dashed curve). (b) $K \ll 1$: jumps among small trapping arcs in a quickly 
evolving potential topography. Solid and dashed contours stand for potential hills 
and wells.
}
\label{fig1}
\end{figure}

\textit{Derivation of FPE from Hamiltonian dynamics for particles}. We consider
model
(\ref{eq2})  where
$\Phi$ is a statistically stationary, spatially homogeneous, isotropic
zero-mean-value stochastic potential with typical amplitude $\Phi_0$, and a 
given two-point, two-time correlation function, with correlation time
 $\tau_c$ and correlation length $\lambda_c$, as considered in
Refs. \cite{ref6,ref12}.
Some results from these references are important for justifying FPE in
this case.  Potential $\Phi$ drives the particles
motion by setting the instantaneous value of their velocity, whose
typical amplitude is  $\Phi_0 / \lambda _c $. The system
chooses among two types of diffusive chaotic dynamics according to the
value of the Kubo number $K = \Phi_0 \tau _c /\lambda _c^2$. The rationale for
this is the
following. If the potential is static ($\tau_c$ infinite),
particles are trapped into potential wells and hills, and possibly make long flights
along
\textquotedblleft{}roads\textquotedblright{} crossing the whole chaotic
domain. These various domains are separated by separatrices
joining nearby hyperbolic points. If  $\tau_c$ is
finite, but large ($K  \gg 1$), the potential topography slowly changes, and the
dynamics
evolves
quasi-adiabatically. Since phase space area inside the
instantaneous closed orbits must be adiabatically
preserved, and since the area of the various domains defined by the
separatrix array fluctuates a lot, orbits must cross the instantaneous
separatrices, and jump this way from one domain to the next one (Fig.
\ref{fig1}(a))
. In the
absence of \textquotedblleft{}roads\textquotedblright{}, which is
almost the case for a gaussian spatial correlation function of the
potential \cite{ref12}, these crossings produce a random walk with step
 $\lambda_c$  and waiting time $\tau_c$; the corresponding diffusion coefficient
is $ D \approx \lambda_c^2 /\tau_c$. The presence of
\textquotedblleft{}roads\textquotedblright{}
modifies this estimate and brings some dependence upon $K$ \cite{ref12}. As a
result,
for $K \gg 1$ diffusion is justified by locality of trapping in phase-space.

When  $K \ll 1$, the particles typically run only along a small arc of length
$\Phi_0 \tau_c / \lambda _c $ of the trapped orbits of the
instantaneous potential during a correlation time. During the next
correlation time they perform a similar motion in a potential
completely uncorrelated with the previous one (Fig. \ref{fig1}(b)). These
uncorrelated
random steps yield a 2D Brownian motion with a QL diffusion coefficient
$ D \approx \Phi_0^2 \tau_c / \lambda _c ^2$. These two limit cases in
$K$ show diffusion is a quite general behavior of
particle transport, even whenever structures are visible in the
electrostatic potential, or whenever non gaussian behaviour is  obtained
for a more limited
statistics, as in Ref. \cite{ref1} whose dynamics, except for isotropy, may be
thought as one realization of that in Ref. \cite{ref12} .

The simple preceding reasoning strongly bears on the stochastic
properties of the potential, and not on the chaotic features of
particle dynamics. A more rigorous picture for the diffusion process of
the $K \ll 1$ regime can be given, which incorporates chaos as an essential
ingredient, but exhibits the paramount importance of potential
randomness. This picture is a translation for dynamics (\ref{eq2})
of that described in Refs. \cite{ref10,ref14} for the dynamics of an electron in
a set of Langmuir waves.
To this end, we consider  $\Phi$ as a sum of
propagating modes
$\varepsilon a_{{\bf{k}},\omega } \cos \left( {{\bf{k}} \cdot {\bf{X}} -
\omega t + \varphi_{{\bf{k}},\omega }} \right)$,
where the  $\varphi_{\bf{k},\omega }$'s are uniformly
distributed random phases, and the spectrum is
isotropic in $\bf{k}$; $\varepsilon$ is put for scaling
purposes. If the particle
does not resonate with the modes in (\ref{eq2}), its dynamics may be
described by perturbation theory in  $\varepsilon$. In
the opposite case, at any given time $t_0$ where the particle has
velocity $\textbf{v}(t_0)$, some modes are resonant
 with the particle, and their action cannot be
described in a perturbative way, but some are not, and still act
perturbatively. These two classes can be defined according to their
resonance mismatch  with the particle
$\rho(t_0)  = \left| {\textbf{k} \cdot \textbf{v}(t_0) - \omega }\right|$
\cite{ref10,ref14,ref15}.
Consequently, the instantaneous particle motion splits into a
perturbative, and thus non chaotic part, and into a non perturbative
chaotic part due to the set $S(\textbf{v}(t_0))$ of modes which are resonant 
enough
with the particle at time $t_0$: $\rho(t_0)$ less than some threshold $\rho_0 
\sim
\varepsilon $. This set evolves
with
time, according to the instantaneous value of $\textbf{v}(t)$.
A suitable choice of $\rho_0 / \varepsilon $ enables to incorporate the set of
modes
driving the chaotic dynamics of the particle over a finite interval
$[t_0 - \Delta t, t_0 + \Delta t]$, where $\Delta t$ is of the order of the
time $\tau_\text{spread} \approx \lambda _c^2 / D $.
Over this time interval the chaotic
particle dynamics is ruled by a reduced Hamiltonian $\Phi _{t_0 } (\textbf{X},t)$,
which is
 $\Phi (\textbf{X},t)$ with the summation restricted over
the modes inside set $S(\textbf{v}(t_0))$. If the dynamics is chaotic,
$\textbf{v}(t)$
changes a lot its direction during its motion, which brings
disconnected sets $S(\textbf{v}(t_n))$, for different
times $t_n, n=0,1,...$. The randomness of the  $\varphi_{\bf{k},\omega }$'s
implies that the dynamics ruled by each of the $\Phi_{t_n }({\bf{X}},t)$'s  are
statistically
independent, which justifies a central limit argument for their
cooperative contribution to the particle motion: the dynamics is
indeed diffusive. \\
The above argument uses the locality of wave-particle resonance
in phase velocity. It also holds to prove the diffusive
behavior for peaked frequency spectrum, or for a single typical outcome of the 
random phases,
if the set of the initial particle velocities
$\textbf{v}_0$ is spread enough for them to be acted upon
by a large number of disconnected
 $S(\textbf{v}_0)$'s. This explains the diffusive behavior found in \cite{ref16}
by averaging over initial particle positions. It should be stressed
that the above locality rationale depends essentially on the random
phases, and that chaos, through the breakup of KAM tori, just brings in
the ability of the motion to be ruled successively by uncorrelated
dynamics. In the absence of random phases no diffusivity can be derived
by the above reasoning. Indeed the diffusion picture was clearly shown
to break down in \cite{ref1}, and for electron dynamics due to Langmuir waves
\cite{ref15}. \\
The existence of random phases can also be used to derive a
rigorous QL estimate for $D$. This derivation is analogous to that for the chaotic
dynamics
due
to Langmuir waves for $K \ll 1$ \cite{ref10}.
Its central argument is that the dynamics depends
slightly on any two phases during a time much larger than
 $\tau_\text{spread}$, which defines the time of
strong sensitivity of the dynamics on the whole set of random phases
\cite{ref14}.
These successive two random-phase arguments for the
$K \ll 1$ case, assumed for simplicity that all phases were uncorrelated, but
some correlation
may be accommodated.
We stress
they do not use at all any loss of memory due to chaotic motion: indeed
differentiable chaotic Hamiltonian dynamics is not hyperbolic.

\textit{Beyond simple Hamiltonian models}. More recently Vlad \textit{et al}.
\cite{ref17}
introduced a spatial inhomogeneity in model (\ref{eq2}),
such that  its r.h.s.  is
multiplied by a growing function of $x_{1}$.
This was meant as a modeling of the increase of the magnetic field
toward the main axis of a fusion machine, but makes the dynamics non
Hamiltonian stricto sensu. On top
of the previous diffusive
behavior (which becomes anisotropic), the new inhomogeneity
brings a \textquotedblleft{}radial\textquotedblright{} drift velocity
 $V$ along  $x_1$ due to the chaotic motion, which corresponds to the dynamic
friction of
FPE. The sign of  $V$ depends on $K$. If $K \ll 1$, since the velocity
increases toward larger $x_1$'s, the
displacement during a correlation time is larger toward the exterior
than toward the interior, which brings an outgoing drift. If $K \gg 1$,
the trapped particles are slower in the inner part of their orbit,
which increases their probability to be there with respect to that to
be in the outer part: this
brings an ingoing drift. Since $D$ now grows with $x_1$, $V = dD/dx_1$ is
impossible for
$K \gg 1$. \\
The previous discussion
can be extended to higher
dimensional systems quite naturally, as do the above arguments of
locality, or the argument of the weak effect of two phases on the
dynamics. The diffusive aspect of higher dof chaotic systems is largely
documented as well
(see \cite{ref17b} and references therein). 
In conclusion we may state that  \textit{depending on the statistics
of interest, the
same dynamical system may be found diffusive or dominated by its L\'{e}vy
flights}. In particular averaging over many random phases may lead to diffusion, while 
averaging only over a limited set of initial conditions for the particles may lead to the L\'{e}vy 
flight picture. Since
fusion experiments generally average over many plasma realizations, and global
confinement
scaling laws even more so, FPE is a highly relevant tool for this field.
It should be noted that in a magnetized toroidal plasma, density $n$ describing
particle transport is the true particle density divided by a growing function
$\tau(x)$ of the local magnetic field (see \cite{ref17c} and references therein).
This brings a slanting of the density profile toward the
outer part of the torus that has nothing to do with a turbulent transport 
phenomenon, in
contrast with
that in Ref. \cite{ref17}, and which corresponds exactly to a waiting time
$\tau(x)$ in Eq.
(\ref{eq3}). This is one more caveat for data analysis. \\
We found that for $K \ll 1$, FPE with a QL
diffusion coefficient is justified by chaotic Hamiltonian dynamics
with random phases, even though structures exist in phase space for one
realization of the phases. This, and the fact that the
QL diffusive modeling of transport is quite efficient \cite{ref18}, suggest
 that $K$ is small in magnetic fusion turbulence. There are reasons for $K$ not
to be large.
First, the usual estimate in fluid mechanics of the correlation time as the eddy
turn-over time
yields $K = 1$ \cite{ref18b}.
Furthermore strong turbulence theory predicts that $K$ is at most of
order 1 (see Eq. 4.34 of \cite{ref19}). However the issue of the typical value of
$K$ is still
unsettled, since an analysis of fluctuation data in the TEXT tokamak indicated
$K$ of order 1, and a poor agreement of QL estimates for impurity transport 
\cite{ref19b}. It
would be
interesting to systematically compute $K$ from experimental or numerical data.
In rotating or stratified fluid turbulence a weak effect of structures
holds as well: cigar-like or pancake-like structures are present, but turbulent
diffusion is correctly modeled by assuming random phases for the
Fourier components of the turbulent fluid \cite{ref20}.
\begin{acknowledgements}
We thank D. del-Castillo-Negrete and G. Spizzo for interactions which
triggered these thoughts, and S. Benkadda, Y. Elskens, X. Garbet, M. Ottaviani, 
and R. Sanchez for
useful
suggestions. This work was supported by the European
Communities under the Contract of Association between Euratom/ENEA.
\end{acknowledgements}

\end{document}